\documentclass[
reprint,
superscriptaddress,
 amsmath,amssymb,
aps,
prapplied,
floatfix
]{revtex4-2}

\usepackage{float}
\usepackage{graphicx}
\usepackage{dcolumn}
\usepackage{bm}
\usepackage{color}

\begin{document}

\preprint{APS/123-QED}

\title{Pure-state photon-pair source with a long coherence time for large-scale quantum information processing}

\author{Bo Li}
\affiliation{Hefei National Laboratory for Physical Sciences at the Microscale and Department of Modern Physics, University of Science and Technology of China, Hefei 230026, China}
\affiliation{Shanghai Branch, CAS Center for Excellence in Quantum Information and Quantum Physics, University of Science and Technology of China, Shanghai 201315, China}
\affiliation{Shanghai Research Center for Quantum Sciences, Shanghai 201315, China}

\author{Yu-Huai Li}
\email{liyuhuai@ustc.edu.cn}
\affiliation{Hefei National Laboratory for Physical Sciences at the Microscale and Department of Modern Physics, University of Science and Technology of China, Hefei 230026, China}
\affiliation{Shanghai Branch, CAS Center for Excellence in Quantum Information and Quantum Physics, University of Science and Technology of China, Shanghai 201315, China}
\affiliation{Shanghai Research Center for Quantum Sciences, Shanghai 201315, China}

\author{Yuan Cao}
\affiliation{Hefei National Laboratory for Physical Sciences at the Microscale and Department of Modern Physics, University of Science and Technology of China, Hefei 230026, China}
\affiliation{Shanghai Branch, CAS Center for Excellence in Quantum Information and Quantum Physics, University of Science and Technology of China, Shanghai 201315, China}
\affiliation{Shanghai Research Center for Quantum Sciences, Shanghai 201315, China}

\author{Juan Yin}
\affiliation{Hefei National Laboratory for Physical Sciences at the Microscale and Department of Modern Physics, University of Science and Technology of China, Hefei 230026, China}
\affiliation{Shanghai Branch, CAS Center for Excellence in Quantum Information and Quantum Physics, University of Science and Technology of China, Shanghai 201315, China}
\affiliation{Shanghai Research Center for Quantum Sciences, Shanghai 201315, China}

\author{Cheng-Zhi Peng}
\affiliation{Hefei National Laboratory for Physical Sciences at the Microscale and Department of Modern Physics, University of Science and Technology of China, Hefei 230026, China}
\affiliation{Shanghai Branch, CAS Center for Excellence in Quantum Information and Quantum Physics, University of Science and Technology of China, Shanghai 201315, China}
\affiliation{Shanghai Research Center for Quantum Sciences, Shanghai 201315, China}

\begin{abstract}

The Hong-Ou-Mandel interference between independent photons plays a pivotal role in the large-scale quantum networks involving distant nodes. Photons need to work in a pure state for indistinguishability to reach high-quality interference. Also, they need to have a sufficiently long coherence time to reduce the time synchronization requirements in practical application. In this paper, we discuss a scheme for generating a pure-state photon-pair source with a long coherence time in periodically poled potassium titanyl phosphate (PPKTP) crystals. By selecting the appropriate pump laser and filter, we could simultaneously eliminate the frequency correlation of the parametric photons while achieving a long coherence time. We experimentally developed this pure-state photon-pair source of 780 nm on PPKTP crystals pumped by a 390 nm pulsed laser. The source provided a coherence time of tens of picoseconds, and it showed to have the potential to be applied in long-distance quantum interference. Furthermore, we experimentally demonstrated the Hong-Ou-Mandel (HOM) interference between two photon sources with visibility exceeding the classical limit.
\end{abstract}
\maketitle

\section{Introduction}

Photon-pair sources are essential in the fields of quantum information processing, which include quantum communication \cite{Ekert:1991kl, PhysRevLett.68.557} quantum computation \cite{Knill:2001is}, and Bell tests \cite{PhysicsPhysiqueFizika.1.195, PhysRevLett.23.880}.
 Complex quantum communication protocols such as teleportation \cite{Bouwmeester1997} and swapping \cite{Pan1998} involve more than two photons, where interference is required between independent photon sources. Thus, independent photon-pair sources should be prepared with high spectral purity.
  To implement these schemes on a large scale, it is crucial for the coherence time of the photon source to be sufficiently long (in the order of picoseconds), which will help mitigate the challenges associated with time synchronization over long distances. Simultaneously, it is essential to thoroughly address other key parameters such as brightness and purity. Moreover, source implementation should consider the flexibility and robustness required for practical applications.

 Nowadays, the most extensively used approach to generate photon pairs is spontaneous parametric down-conversion (SPDC) in nonlinear crystals \cite{PhysRevLett.25.84}. 
The pulse duration, material, thickness, and cut angle of a nonlinear medium, or the length and period of a periodically poled nonlinear medium should be carefully considered to deal with the group velocity match (GVM) issue \cite{Grice1997}. 
In conventional multiphoton setups, the common used crystal is $\beta$-barium borate (BBO) or bismuth borate (BiBO) crystals with a thickness of a few millimeters, resulting in a photon bandwidth of 3 (or 8) nm \cite{Yao2012,Wang:2016dk, PhysRevLett.121.250505}. This configuration, operated with a high-power ($\sim$ Watt-level) and femtosecond ($\sim$ 100 fs) pump laser, lacks flexibility when applied in practical field experiments or a power-starved scenarios such as in a satellites.   
With $\sim$ 100 fs of pump pulse \cite{Yao2012}, the synchronization precision should be at the level of ten femtoseconds between remote parties, which reaches the state-of-the-art technology that involves optical clocks \cite{RevModPhys.87.637}, cavity-stabilized
lasers \cite{Kessler2012}, and frequency comb assisted time transfer \cite{PhysRevX.6.021016}, which are not currently practical.
For practical applications, it is advisable to ensure a long coherence time of the source comparable to the practical time synchronization achieved through commercial atomic clocks or oven-protected crystal oscillators \cite{PhysRevLett.125.260503}. Typically, time synchronization systems utilizing these methods yield a precision of tens of picoseconds.

By taking advantage of the quasi-phase-matching (QPM), periodically poled nonlinear crystals can provide a flexible approach for the almost arbitrary phase-matching angles and wavelengths \cite{Fedrizzi:07}. Thus, with a collineation configuration, the crystal thickness can be fairly long for a high generation rate of photon pairs without unwanted walk-off. QPM-SPDC sources have a variety of applications, such as space-borne \cite{Yin2017, Yin2020, Yin2017PRL, Tang2016} and air-borne payloads \cite{10.1093/nsr/nwz227, PhysRevLett.126.020503} in long-distance quantum communication, photonic boson sampling \cite{Zhong2021} in quantum computation.
A series of great efforts have been made concerning the telecom photon pairs generated from PPKTP \cite{JRB2014, Jin:2015bu} for multiphoton applications.
When preparing degenerate parametric photons at around 1560 nm  in type-II PPKTP crystals, the GVM condition can intrinsically be satisfied. Thus, the time and frequency correlations between the signal and idler photons are naturally eliminated even without any filters. However, due to the wavelength increase, the emission rate of photon pairs at telecom band is decreased compared with the configuration of visible parametric photons \cite{PhysRevA.77.043834}.
PPKTP-based photon-pair sources having a wavelength of 780 nm, which are suitable for non-classical interference, have been reported \cite{Scheidl:2014gs}. However, the configuration employs a short crystal (1 mm) and ultrafast pump pulses (150 fs), leading to
a short coherence time of $\sim$ 1 ps.

In this paper, by carefully examining the GVM condition, we find that filtered SPDC sources can still be beneficial for application in the field experiment with an acceptable loss. 
We show that a 780-nm photon-pair source with a bandwidth of 30 pm can be constructed by carefully selecting the pump laser, nonlinear crystal parameters, and proper filters on the parametric photons.
This implies that a coherence time of tens of picoseconds and a pair generation rate at the order of $\sim 2\times10^5$ pairs/mW/s could be reached.
With such a configuration, a 30-mW pump laser with a repetition rate of 76 MHz was sufficient to produce $\sim$ 0.1 photon pairs per pulse, which lower the requirement of high-power pump condition. 
Furthermore, a proof-of-principle experiment of the Hong-Ou-Mandel (HOM) interference \cite{1987PhRvL..59.2044H} between two heralded single-photon sources was performed to demonstrate the feasibility of the proposed method in the multiphoton applications. 

\section{Design of the photon source}

When developing a photon-pair source, it is crucial to consider the photon spectral emission and phase-matching conditions. The photon yield depends on several factors, including the wavelength employed, the crystal's length, and the focusing parameters in bulk crystals. The phase-matching conditions affect the spectral distribution of the generated photons. Understanding these conditions is essential for establishing the ideal filtering approach and other specific applications.

\subsection{General concepts}
\subsubsection{Spectral emission}

According to the theory of the SPDC process in the bulk crystals, in the case of a collinear emission with the degenerate wavelength, the spectral generation rate of parametric photons can be written as follows \cite{2007OExpr..15.7479F}:

\begin{equation}
 \frac{d P_{s}}{P_p \cdot d\omega_{s}} = \frac{\hbar d^2 {L} \omega_s^4 } {2\pi c^4 \varepsilon_{0} n_{p}^{2}} f\left (\lambda_{s}\right) = \frac{\hbar d^2 } {2\pi c^4 \varepsilon_{0} } f\left (\lambda_{s}\right) \cdot \frac{1} {n_{p}^{2}} \cdot {L} \omega_s^4
\label{Ps}
\end{equation}
where $\omega_s$, $\omega_i$ and $\omega_p$ are the angular frequencies of the signal, idler and pump photons, respectively.
$P_s$ is the signal photon power integrated over all emission angle.
$d$ is the effective nonlinear coefficient, $L$ is the crystal length, $\omega$ is the angular frequency, $c$ is the speed of light in vacuum, $\varepsilon_0$ is the permittivity of vacuum, $n_p$ is the refractive index of the pump laser in the crystal, and $P_p$ is the pump power.
$f (\lambda_s )$ is a geometry related function, and it be regarded as a constant in this paper. The first term $\frac{\hbar d^2 } {2\pi c^4 \varepsilon_{0} } f\left (\lambda_{s}\right)$ is a constant. The value of the second term $\frac{1} {n_{p}^{2}}$ depends on $\omega_p$.
For a polarized photon transmission along the y-axis in a PPKTP crystal, the refractive index varies from 1.844 to 1.727 for a wavelength range of 0.4-2.0 $\mu$m \cite{Kato:02}. This variation is negligible, as it is less than 7\%. As indicated in the last term, the spectral brightness is directly proportional to the crystal length and the fourth power of $\omega_s$. Therefore, increasing the spectral brightness may be achieved by using a longer crystal and shorter wavelength. In our scheme, it should be noted that filter will not impact the spectral emission. Thus, maximizing the initial spectral emission rate is imperative to increase the available photons with a narrow filter.

\subsubsection{Spectral correlation}
This part details the spectral correlation effect of the general SPDC process.
Theoretically, the wave function of the parametric photons can be expressed as follows \cite{PhysRevA.64.063815}
\begin{equation}
  \vert \psi\rangle=N\int \int d\omega_s d\omega_i f (\omega_s,\omega_i) \hat{a}^{\dagger}_s (\omega_s)  \hat{a}^{\dagger}_i (\omega_i) \vert vac \rangle,
\end{equation}
where $\hat{a}_s^{\dagger} (\omega_s)$ and $\hat{a}_i^{\dagger} (\omega_i)$ are the photon creation operators for the signal and idler beams, respectively, $vac$ is vacuum state, and $N$ is a normalization constant.
$f(\omega_s,\omega_i)$ is the joint spectral amplitude (JSA), which represents the spectral correlation between the signal and idler photons, and it can be expressed as the product of the pump envelope function $ \alpha (\omega_s, \omega_i) $ and QPM function $ \phi (\omega_s, \omega_i) $:
\begin{equation}
\label{equ:jsa}
  f (\omega_s,\omega_i)=\alpha (\omega_s+\omega_i)\phi (\omega_s,\omega_i).
\end{equation}

In the case of pumping by a Gaussian line shaped laser with a central frequency of $\overline \omega_p$ and a bandwidth of $\sigma_p$, the pump envelope function can be written as follows:
\begin{equation}
\label{equ:pump}
\alpha (\omega_s+\omega_i)\propto exp[-\frac{ (\omega_s+\omega_i-\overline \omega_p)^2}{2\sigma_p^2}].
\end{equation}
The phase-matching function is given by the following:
\begin{equation}
\label{equ:pha}
\phi (\omega_s,\omega_i) = sinc[\frac{L}{2}\Delta k],	
\end{equation}
where $\Delta k=k_p-k_s-k_i-m\frac{2\pi}{\Lambda}$, $L$ is the crystal length, $k$ is the wavenumber, $\Lambda$ is the poling period, and $m$ is the order of QPM.

The spectral correlation of photon pairs is determined by the joint spectral intensity (JSI) $S(\omega_s,\omega_i)=|f(\omega_s,\omega_i)|^2$, which is the product of the pump envelope intensity (PEI) $\vert \alpha (\omega_s+\omega_i)\vert^2$ and the phase-matching intensity (PMI) $\vert\phi (\omega_s,\omega_i)\vert^2$. When there is a strong correlation between the photons, the initial purity of the generated photon pair will be low. One way to quantify the purity of the photon pair is by using the Schmidt decomposition on the PSI, see Appendix A.

\subsubsection{Filtering method}

Various technical approaches are available to select the frequency mode, thus enhance the purity, one of which is the cavity-enhanced SPDC technology \cite{baoGenerationNarrowBandPolarizationEntangled2008,  scholzSinglemodeOperationHighbrightness2009, pomaricoEngineeringIntegratedPure2012, monteiroNarrowbandPhotonPair2014, luoDirectGenerationGenuine2015, rambachSubmegahertzLinewidthSingle2016, niizekiUltrabrightNarrowbandTelecom2018, moqanakiNovelSinglemodeNarrowband2019, slatteryBackgroundReviewCavityEnhanced2019}. In this technique, the photon has a typical bandwidth of several hundred MHz, making it well-suited for quantum storage. There are studies focused on developing it towards long-range quantum communication \cite{monteiroNarrowbandPhotonPair2014, niizekiUltrabrightNarrowbandTelecom2018}. However, manipulating cavity-enhanced SPDC is relatively complex, as it requires handling issues such as cavity design, stability, compensation, loss control, and mode selection. The current research limit is the source brightness and efficiency, making it less effective for practical long-range quantum communication.

In comparison, using a narrow bandwidth filter to tailor the broadband SPDC photons is a straightforward and flexible way to increase photon purity and coherence length. Supposing that the interference filter is a Gaussian type filter with a bandwidth of $\sigma$, then its amplitude function is in the form of the following:
\begin{equation}
  g (\Delta\omega)= e^{-{\Delta\omega}^2/2\sigma^2 }.
\end{equation}
By introducing this to both signal and idler photons, the filtered JSA can be represented as follows:
\begin{equation}
f^\prime (\omega_s,\omega_i)=g (\omega_s - \overline\omega_s)g (\omega_i - \overline\omega_i)f (\omega_s,\omega_i).
\end{equation}
Similarly, the purity can be obtained by Schmidt decomposition, for an optimal filtering strategy.

\subsection{Detailed design}

Below we use a long PPKTP crystal as an example to analysis how to design the source.

In engineering a photon-pair source for practical multiphoton use, achieving high purity is crucial. This can be accomplished either by manipulating the natural SPDC condition to meet specific requirements, or by filtering the spectrum afterwards. A proper filter not only increases purity, but also enhances coherence time. However, the use of filters may result in photon loss, which can be mitigated by selecting high-performance periodically poled crystals that possess a high initial photon flux. 

The factor that affect spectral brightness are represented in Eq. \ref{Ps}. To achieve a high photon flux, a type-II PPKTP crystal with a length of 30 mm and a poling period of 7.825 $\mu$m is used. The crystal was designed for degenerate emission with a pump laser of 390 nm, and it is crucial that the pump beam is focused onto the center of the crystal with an appropriate waist.
A loose focus increase the coupling efficiency, while a strong focus can increase the pair emission rate. A trade-off between coupling efficiency and emission rate is necessary depending on specific applications \cite{PhysRevLett.121.080404, Zhong2021}. For further detailed analysis, refer to \cite{Fedrizzi:07,PhysRevA.77.043834}.

\begin{figure}[htbp]
  \centering
  \includegraphics[width=8.6cm]{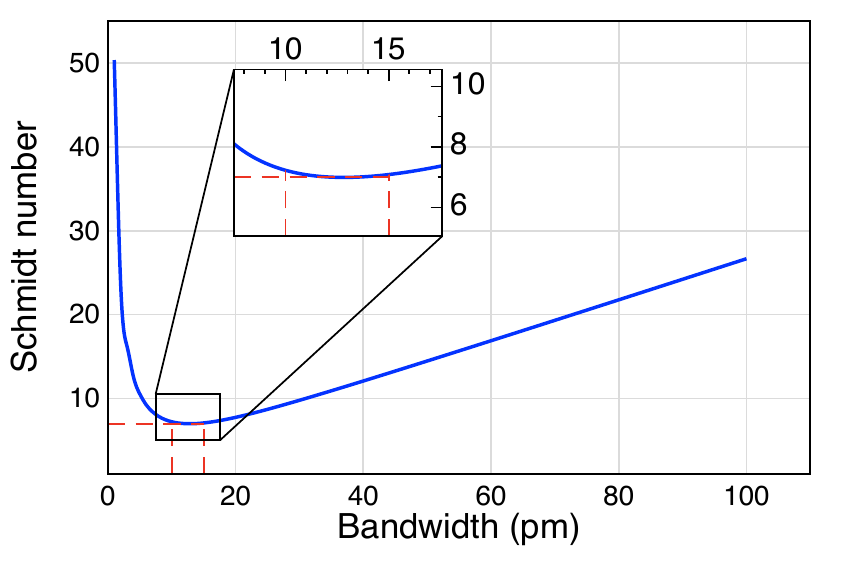}
  \caption{Schmidt numbers with different pump bandwidths. A lower Schmidt number implies a better natural purity of the generated photon pairs, and a less discard fraction in filtering to reach a high purity. We showed the interval with the lowest Schmidt number of $\sim$ 7 (enlarged in the figure).}  
  \label{fig:Schmidt number}
\end{figure}

The pump laser bandwidth is another important factor to consider. The bandwidth has less effect on the absolute photon emission, but it greatly affect the distribution of the JSI, thus the final filter selection. In our scheme, we aim to preserve as many photons as possible, so an optimal bandwidth around 10-15 pm is chosen at the lowest Schmidt number, as shown in Fig. \ref{fig:Schmidt number}.
A lower Schmidt number represents a less mode number in the initial spectrum, thus reducing the filtering loss. From this, the optimized bandwidth of the pump laser could be obtained.  

\begin{figure}[htbp]
\centering
\includegraphics[width=8.6cm]{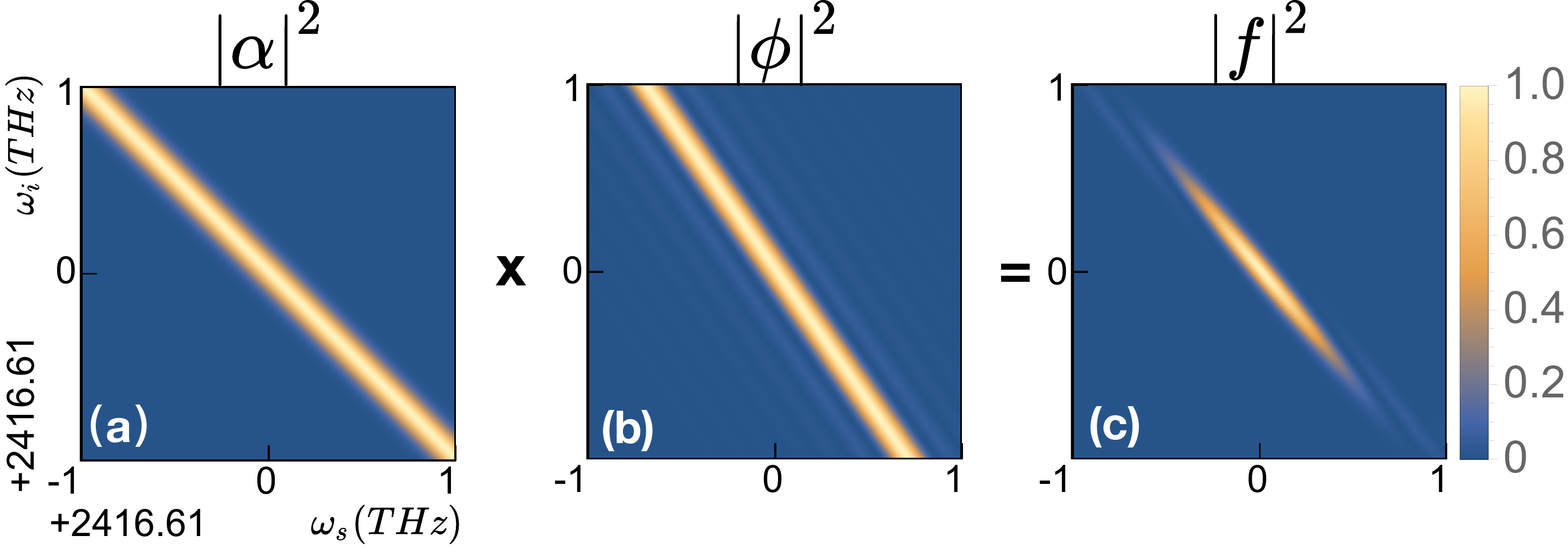}
\caption{Example of the (a) pump envelope intensity, (b) phase-matching intensity, and (c) joint spectral intensity for a specific configuration: type-II PPKTP with a length of 30 mm, a pump laser with a central wavelength of 390 nm and a full width at half maximum (FWHM) of 15 pm. A strong anti-correlation between the frequencies of the signal and idler photons can be seen in (c).}
\label{fig-jsi}
\end{figure}

The corresponding JSI distribution is shown in Fig. \ref{fig-jsi}. The PEI's slope is fixed to $135^{\circ}$ due to energy conservation, and its width is proportional to the bandwidth of the pump laser $\sigma_p$. The width of PMI is inversely proportional to the length of the PPKTP crystal $L$, while the slope is highly dependent on the wavelengths of the pump, signal, and idler photons. for the type-II SPDC process of $390 ~nm \to 780 ~nm + 780 ~nm$, the slope of PMI is $124.8^{\circ}$, which is close to the slope of PEI, resulting in a highly correlated JSI. This type of photon source is intrinsically spectral correlated. Thus, it cannot be directly applied in multiphoton engineering without first eliminating the correlation to achieve high purity. 

\begin{figure}[htbp]
  \centering
  \includegraphics[width=8.6cm]{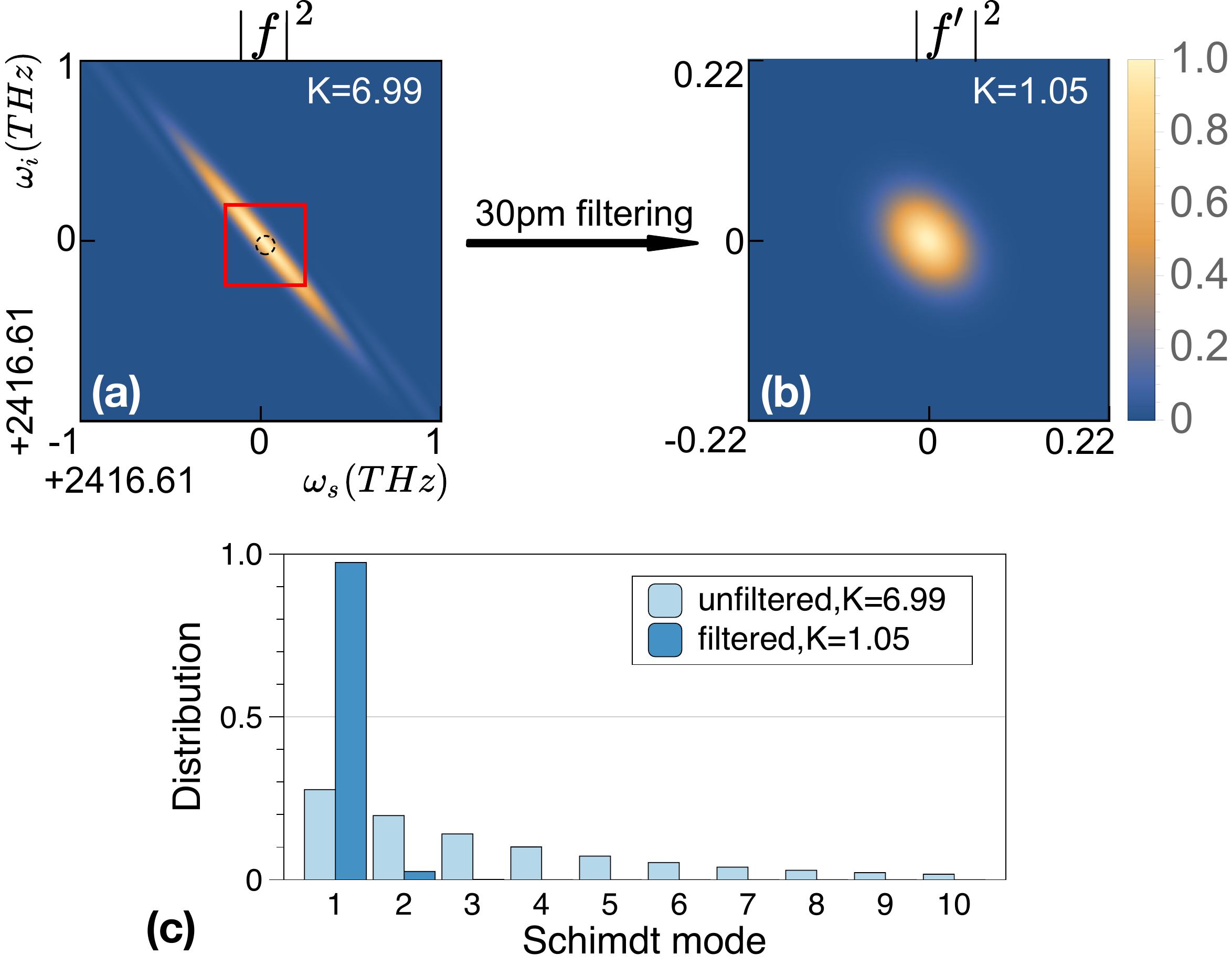}
  \caption{Frequency correlation and Schmidt mode for the signal and idler photons. (a) The original JSI is highly correlated in frequency, with a Schmidt number of K = 6.99. (b) By filtering both the signal and idler photons with a 30 pm filter, the frequency correlation is almost completely removed, where the Schmidt number K = 1.05. (c) Comparison of the Schmidt mode distributions for the original and the filtered JSI.}
  \label{fig:spectralBrightness}
\end{figure}

To further improve the Schmidt number to approach 1, a filtering method is proposed to be applied to the signal and idler photons symmetrically. As shown in Fig. \ref{fig:spectralBrightness}, by applying a filter with an appropriate bandwidth ($\sim$30 pm), the Schmidt number mode can decrease to 1.05. The Schmidt mode distributions for the filtered and the original frequency correlations are shown in Fig. \ref{fig:spectralBrightness}(c).

\begin{figure}[htbp]
  \centering
  \includegraphics[width=8.6cm]{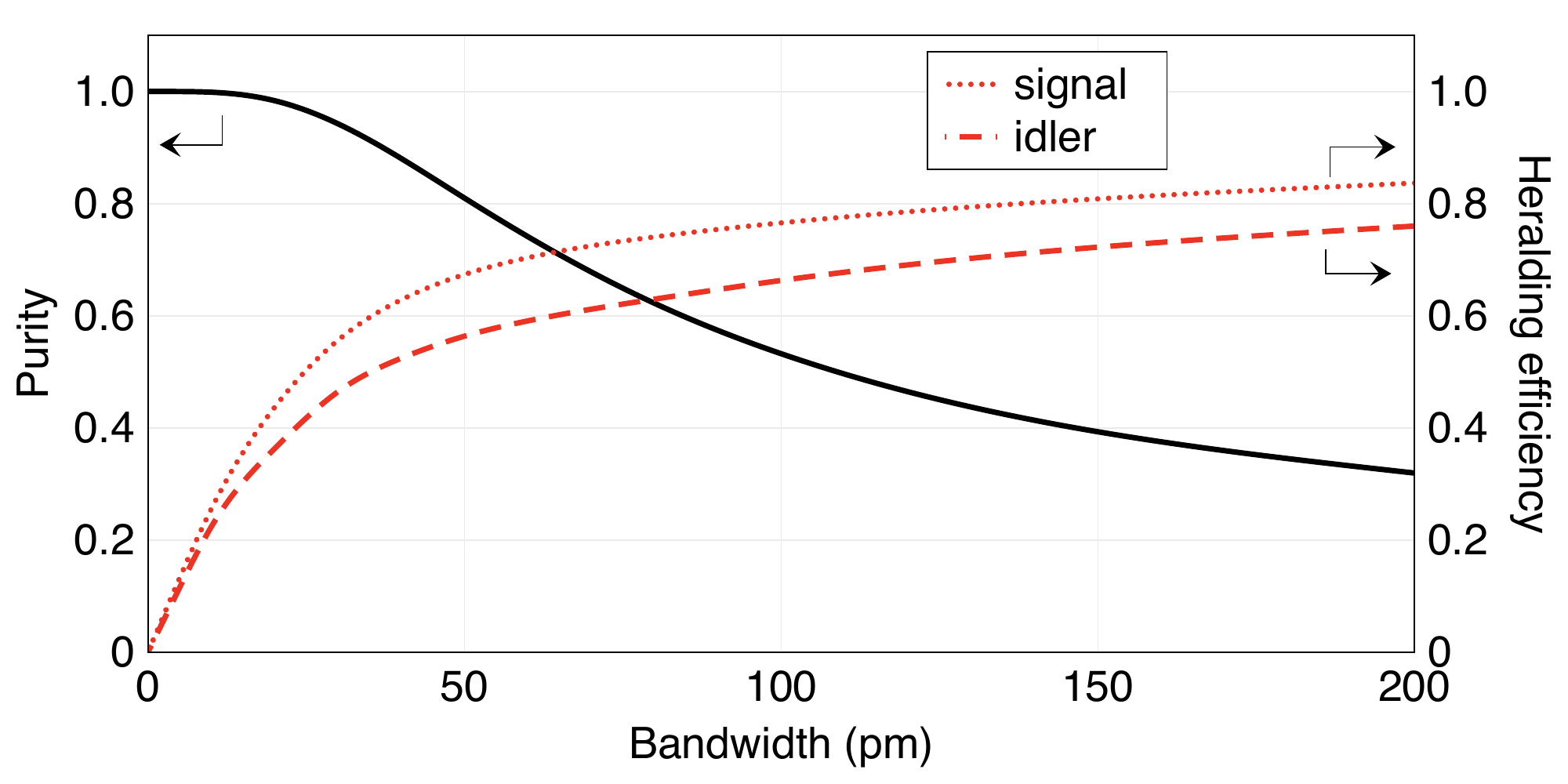}
  \caption{
  The relationship between filter bandwidth and spectral purity, and the heralding efficiency for signal and idler photons. The filter is applied symmetrically to both signal and idler photons. As the filter becomes narrower, the purity (indicated by a solid black line) increases gradually from a low level to 1. However, simultaneously, the heralding efficiency (indicated by red dashed lines) decreases towards 0. Therefore, one must consider the trade-off between the purity and the heralding efficiency. For our application, an optimal filter bandwidth of roughly 30 pm is recommended, as this allows the purity to reach 0.95.
  }

  \label{fig:visibility}
\end{figure}

According to the study by Meyer-Scott et al. \cite{meyer-scottLimitsHeraldingEfficiencies2017}, there is a trade-off between purity and heralding efficiency, which can be characterized by examining the JSI distribution. The filter heralding efficiency is defined as the ratio of the probability that both photons pass through their respective filters to the probability that each individual photon passes through its filter, denoted $\eta_{f,\{s,i\}}=\Gamma_{both}/\Gamma_{\{s,i\}}$.
Fig. \ref{fig:visibility} illustrates this trade-off for parametric photons.  
When the filter bandwidth is approximately 30 pm, the source achieves a high purity of $P=0.95$. However, the corresponding filter heralding efficiencies for the signal and idler photon reduce to $\eta_{f,s}=55.6\%$ and $\eta_{f,i}=46.5\%$, respectively, with an average of 50.8\%.

\begin{table}[htbp]
\caption{\label{tab:table1}The optimal parameters for filtered SPDC.}
\begin{ruledtabular}
\begin{tabular}{lccccccccc}
Parameters & Without filter & filter\footnotemark[1]  \\ \hline
Pump power (mW)&  30  & 30 \\ 
$\Delta\lambda_{p}$ (nm) & 0.015 & 0.015\\
$\Delta\lambda_{\{s,i\}}$(nm)& 0.21/0.26   & 0.03/0.03   \\
Generation rate (s$^{-1}$)  &93.9M & 6.0 M  \\
Coincidence rate (s$^{-1}$) & 3.75 M  &  0.237 M    \\
Heralding efficiency &  20.0\%  &    $\sim$10.2\%  \\
$\overline{n}$\footnotemark[2]  &     &  0.079    \\
Purity & 0.14 & 0.95  \\
\end{tabular}
\end{ruledtabular}
\footnotetext[1]{Ideal Gaussian filters are symmetrically applied on both signal and idler photons.}
\footnotetext[2]{Average photon number per pulse with the repetition set as 76 MHz}
\end{table}

Table \ref{tab:table1} presents one optimal parameters of the filtered SPDC process. The first column presents data measured using 30 mm bulk PPKTP crystals. The typical generation rate 3.1 M pairs/mW/s and the bandwidths of idler and signal photons are 0.21 and 0.26 nm, respectively. The heralding efficiency is about 20\%, which includes the single-mode coupling efficiency of about 40\% and a typical Si detector efficiency of about 50\%. The second column presents the predicted data with an ideal 30 pm filter employed symmetrically on both photons.  The filtered SPDC designed in this paper requires a pump power of approximately 30 mW to produce an average photon number of $\overline{n}\approx 0.08 $. The two-fold coincidence rate has the potential to reach 237,000 counts per second. Our analysis indicate that filtered SPDC is still feasible in creating high-purity photon sources, which could have important implications for quantum information processing.

It's worth noting that waveguides are emerging as a practical alternative to bulk crystals in the field of photonics \cite{halderHighCoherencePhoton2008,levineHeraldedPurestateSinglephoton2010}. By utilizing long waveguides instead of bulk crystals, photon emission in SPDC can be enhanced by one order of magnitude or more. This enhancement is attributed to waveguides offering a higher confinement of the pump field, which leads to an increased photon-pair emission rate. Advancements in waveguide technology are anticipated to address concerns regarding the uniformity of long crystal manufacturing, the efficiency of fiber coupling, and integration with free-space optics, which ultimately paves the way for more versatile applications.

In recent years, novel quantum photon sources, such as semiconductor quantum dots \cite{wangOnDemandSemiconductorSource2019} and atomic ensembles \cite{parkPolarizationEntangledPhotonsWarm2019,parkDirectGenerationPolarizationentangled2021,parkGenerationBrightFourphoton2022}, have emerged as promising way to reach high brightness, high purity, and long coherence time, which are essential for many quantum applications.  
The semiconductor chip construction is complex and requires a low-temperature bath. 
While semiconductor quantum dots offer a near-perfect solution to the problem of low average photon number per excitation pulse, they are still in early development stages and have not yet surpassed mature SPDC technology in terms of flexibility and simplicity.

\section{Experimental implementation}

\begin{figure*}[htbp]
\centering
\includegraphics[width=14cm]{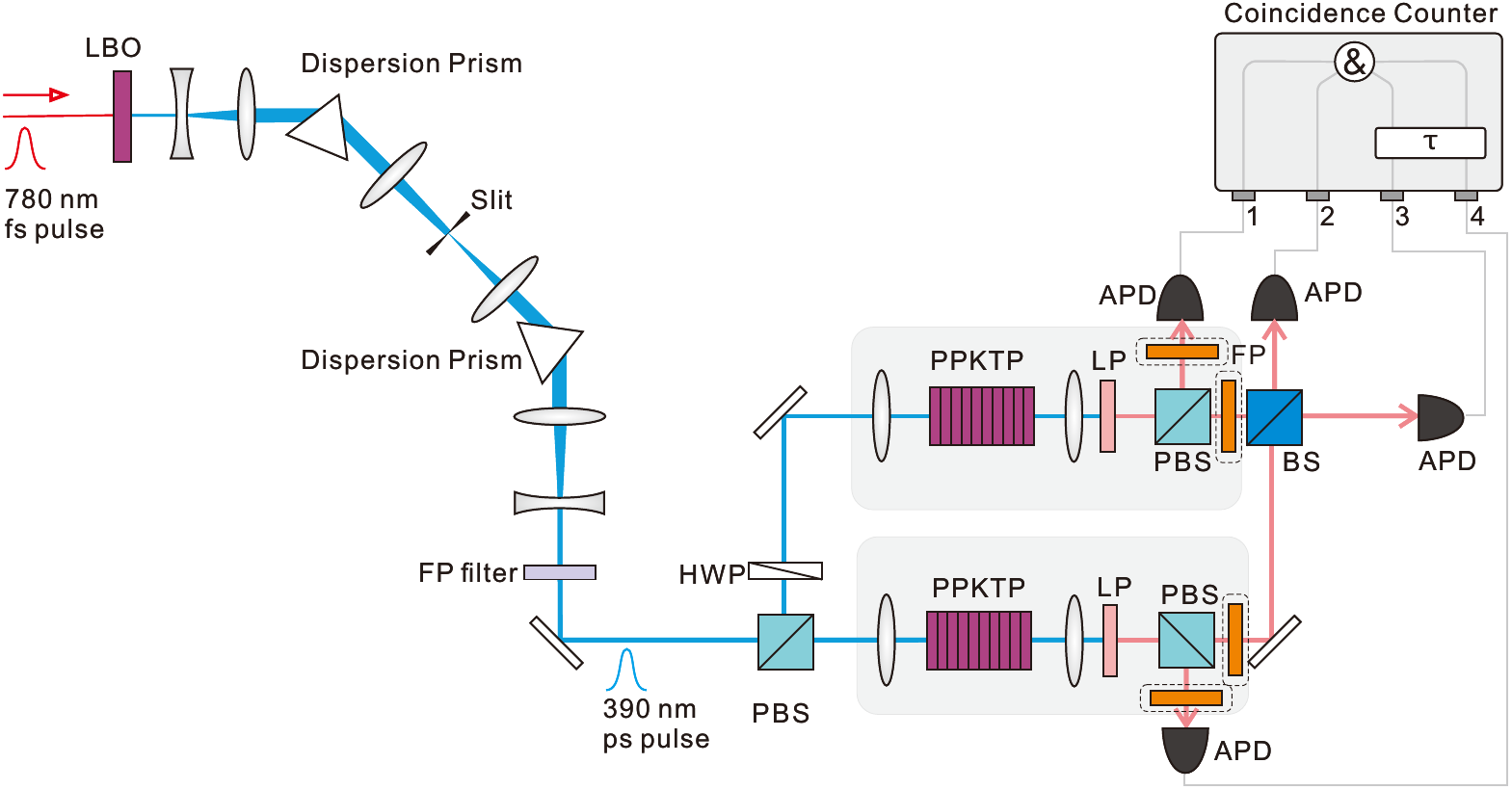}
\caption{Experimental demonstration of the HOM interference. LBO, 1-mm thick lithium triborate; PBS, polarizing beam splitter; HWP, half-wave plate; LP, long-pass filter to block 390 nm; BS, 50:50 nonpolarization beam splitter; APD, Si avalanche photodiodes with a detection efficiency of $\sim$ 0.5. $\tau$ is the digital delay applied to photons 3 and 4.}
\label{fig:setup}
\end{figure*}

Based on the above analysis, we experimentally realized a frequency-uncorrelated photon-pair source. We employed a type-II PPKTP crystal with 1 mm $\times$ 2 mm $\times$ 30 mm for SPDC.
The incident laser was focused with a waist diameter of $\sim$ 50 $\mu$m at the center of the PPKTP crystal. For the polarization, the pump and signal photons were adjusted to be along the crystal y-axis, while the idler photons were adjusted to be along the z-axis. Thus, the signal and idler photons could be separated by a polarized beam splitter, then coupled into single-mode fibers, and detected by Si avalanche photodiodes (APDs), respectively. . 

To perform the experiment, a pump laser with a pulse width of $\sim$15 picoseconds is required. Unfortunately, we did not have access to such a laser. To address this requirement, we modified an fs laser into a transform-limited ps laser using a filtering approach. 
As illustrated on the left side of Fig. \ref{fig:setup}, the fs pulsed laser with a wavelength of 780 nm was frequency-doubled at a 1-mm-thick lithium triborate (LBO) crystal to generate a pulsed laser of 390 nm.
The 780-nm fs pulse came from a Ti: sapphire laser with a repetition rate of 76 MHz. The 390 nm laser has a FWHM of 1.6 nm and average power of 1.6 W. 
An air-spaced Fabry–P\'{e}rot (FP) cavity with a finesse of 20 and an FWHM of 15 pm was selected as the narrow filter. The free spectral range of the FP filter was $\sim$ 1.6 nm. Thus, coarse filtering was necessary to cover the range. Another consideration was the high-power fs laser; we employed a combination of a pair of dispersing prisms with a dispersion coefficient d$\theta$ / d$\lambda$ = 0.65 mrad / nm and a slit as the coarse filter. Before entering the first prism, the pump beam was expanded into a diameter of $\sim$ 5 mm and a divergence angle of $\sim$ 100 $\mu$rad by the lens. An adjustable slit was placed at the focus plane to block out any unwanted spectra. Afterward, the beam was reshaped, collimated, and then passed through the FP cavity for fine filtering. According to the time-bandwidth product limit, the final pump pulse corresponded to a pulse width of $\sim$15 ps. The remaining available pump power was $\sim$800 $\mu$W in total.

To improve the spectral purity of parametric photons, Fabry-Perot (FP) cavities with a full width at half maximum (FWHM) of 30 pm and a finesse of 20 were employed for filter purpose.
To assess the purity of the parametric photons, the second-order autocorrelation function $g^{(2)}{(t)}$ was measured in a Hanbury Brown and Twiss-like configuration \cite{brown1956correlation}, where photon coincidence was detected after a balanced beam splitter. The purity is related to the second-order autocorrelation function through the expression $g^{(2)}(0)=1+P$, where $P$ denotes the purity. For an ideal decorrelated state, $g^{(2)}(0) = 2$ with $P=1$, while a strongly correlated state exhibits Poissonian statistics with $g^{(2)}(0) = 1$ and $P=0$.
In this study, numerical estimates indicated that the spectral purity was $P_s = 0.7$ and $P_i = 0.9$ for perfectly filtering the signal and idler photons, respectively. 
The measured value of $g^{(2)}(0)$ was $1.01\pm0.02$ for unfiltered photons, indicating a very low purity. When the filter was applied to the idler photons, $g^{(2)}(0)$ increased to $1.72\pm0.02$, indicating a purity of approximately $P_i\sim$0.72.
To further enhance the purity, another filter could be employed on the signal photons. It is estimated that the purity could reach as high as 0.95 in theory when both photons are filtered.

After the filtering process, the observed coincidence count rate was $1.5\times10^3$ pairs/mW/s, corresponding to an estimated photon pair generating rate of $2\times10^5$ pairs/mW/s ($\sim 0.003$ pairs/mW per pulse).
For most multiphoton experiments and applications, considering the signal-to-noise ratio,  the optimized generating rate of the parametric photon-pair is less than $ 0.1$ per pulse. 
Hence, a 30-mW pump laser is sufficient.
The measured heralding efficiency of each parametric photon was $\sim 4.3\%$, which is lower than the predicted value listed in Table \ref{tab:table1}. The extra loss can be attributed to the spectral line shape of the FP cavity, the transmittance, and the distortion of the wavefront by the cavity. 
By improving the filtering performance and increasing the pump power, a higher coincidence count rate can be expected in the future.

HOM interference is one direct way to verify the purity of SPDC photon pair source. The theory behind HOM interference can be found in Appendix B.  It should be noted that spectral correlation is associated with photon pairs joint distribution. The HOM interference is heralded by the idler photons.
It is essential to recognize that spectral correlation is associated with the joint distribution of photon pairs. HOM interference is heralded by the idler photons, filtering them can improve spectral purity, leading to and enhancement in the visibility of the signal photon's HOM interference.
We built a HOM interference configuration by preparing and combining two photon-pair sources, as shown on the right side of Fig. \ref{fig:setup}.
The idler photons worked as triggers, and the signal photons were combined on a beam splitter for interference.
A thermoelectric cooler controlled the temperature of each PPKTP crystal to ensure that the signal (idler) photons have the same central wavelength. 
All the arrival times of the photons were recorded by a time-digital converter (TDC) to generate the four-fold coincidence events. When the signal photons were indistinguishable in all degrees of freedom, they bunched and left the same port of BS. Thus, the four-fold coincidence rate was ideally dorpped to near zero ideally.

\begin{figure}[htbp]
  \centering
  \includegraphics[width=8.6cm]{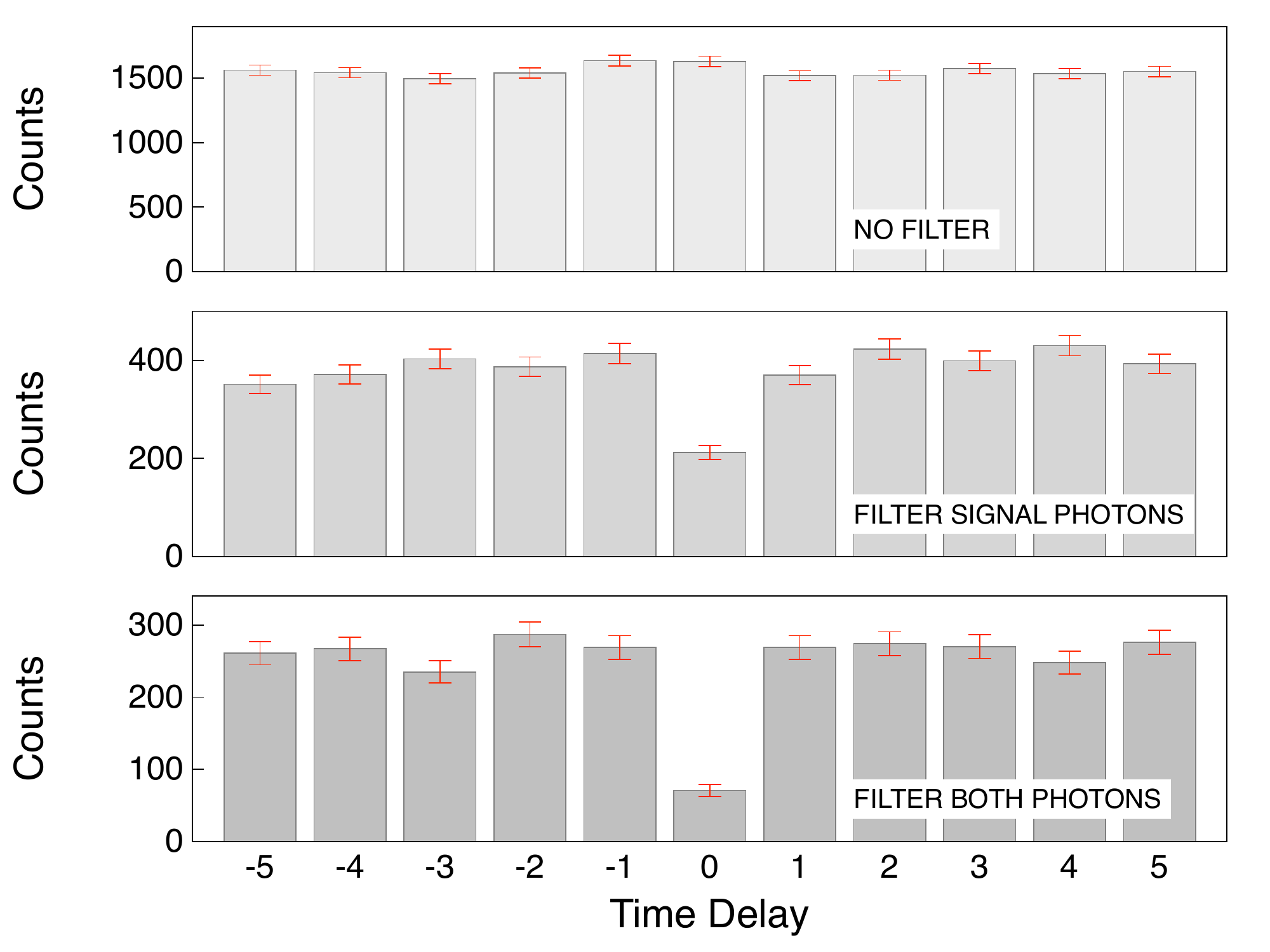}
  \caption{Four-fold coincidence counts with different time delays. The x-axis represents the number of delayed pulse intervals. Since the repetition rate of the pump laser was 76 MHz, the interval of each pulse was 13.16 ns. Without the purification by filters, the visibility of the HOM interference is almost 0. When both signal and idler photons were filtered by FP with an FWHM of 30 pm, the visibility increased to 73.5\% that exceeding the classical limit.}
  \label{fig:result}
\end{figure}

Digital delay scanning was employed to evaluate the visibility of the HOM interference in the experiment.
Before calculating the four-fold coincidence rate, a digital time delay $\tau$ was applied to two of the detectors, as shown in Fig. \ref{fig:setup}. The digital time delay was set in a sequence of the integer multiple of the pulse interval.
The four-fold coincidence counts were displayed with the different digital time delays $\tau$, and the corresponding experimental results are presented in Fig. \ref{fig:result}. When unfiltered, the HOM interference results in low visibility close to 0. We only filtered the signal photons by FP and left the trigger photons unfiltered. Thus, the central bin was relatively dropped, and the visibility increased to 46.01 $\pm$ 2.95\%. 
Further, when the photons of the four paths were all filtered by the FP filters, the visibility reached 73.48 $\pm$ 2.38\%, exceeding the classical limit 0.5 \cite{Bouchard2021,Paul1986}, which provides solid evidence of the nonclassical interference between the two photon-pair sources.

\section{Conclusion}
In summary, using a filtering method, we proposed an approach for building a frequency-uncorrelated photon-pair source at 780 nm. Then, we experimentally realized this approach and showed its potential in multiphoton applications by verifying the HOM interference. The corresponding coherence time of the photon source was as long as tens of picoseconds, which is suitable for large-scale multiphoton applications in which clocks are required to be maintained and independently synchronized in remote parties.
In the future, by enhancing our configuration efficiency and updating the pump laser, we hope to develop the photon-pair source into a quantum communication application for long-distance channels. 

Furthermore, the filtered SPDC technique presented in this study can be applied to research within telecom frequency regime. The spectral brightness of SPDC is known to decrease significantly as wavelength increases. To overcome this limitation and achieve high brightness and long coherence length, ultrabright SPDC sources based on periodically poled lithium niobate (PPLN) waveguides can be explored using the approach discussed in this paper. Furthermore, commercial telecom dense wavelength division multiplexing (DWDM) technologies can be combined for enhanced versatility in constructing a multiplexed photon source, as demonstrated in previous studies \cite{wengerowskyEntanglementbasedWavelengthmultiplexedQuantum2018,xueUltrabrightMultiplexedEnergyTimeEntangled2021}.

\section*{Acknowledgements}

This work was supported by the National Key R\&D Program of China (Grants No. 2017YFA0303900), the National Natural Science Foundation of China (Grants No. U1738201, No. 11904358, No. 61625503, and No. 11822409), the Chinese Academy of Sciences (CAS), Shanghai Municipal Science and Technology Major Project (Grant No. 2019SHZDZX01), and Anhui Initiative in Quantum Information Technologies. Y. C. was supported by the Youth Innovation Promotion Association of CAS (under Grant No. 2018492).

\appendix

\section{Schmidt decomposition}

The purity of the photons is defined as $P_s=Tr (\hat{\rho}_s^2)$, where $\hat{\rho}_s=Tr_i (\left| \Psi \right\rangle\left\langle \Psi \right | )$ is the reduced density operator of the signal \cite{Mosley2008}. The factorability of JSA determines the purity.
When JSA results in a strong correlation, the Schmidt decomposition of the JSA can be used to quantify the purity property \cite{gerry2005introductory}. The Schmidt decomposition can express the joint function into the terms of a set of orthogonal basis. The JSA can be expressed as follows:
\begin{equation}
  f (\omega_s,\omega_i ) = \sum_k \sqrt{\lambda_n}u_k (\omega_{s}) v_k (\omega_{i}) ,
\end{equation}
where $u_k (\omega_{s})$ and $v_k (\omega_{i})$ are two orthogonal basis sets of spectral functions, known as Schmidt modes. $\lambda_n$ is the weight of the Schmidt mode, and  $\sum_n \lambda_n=1$.
The Schmidt number K is defined as follows:
\begin{equation}
\label{equ:K}
  K=\frac{1}{\sum_n \lambda_n^2}.
\end{equation}
K is an indicator of entanglement. It represents the number of the Schmidt modes existing in the two-photon states. The spectral purity is the inverse of the Schmidt number K:
\begin{equation}
	P=P_s=P_i=\frac{1}{K}.
\end{equation} 

For a two-photon state $\vert \psi\rangle$ not correlated in the spectrum, the JSA can be written into a completely factorizable form:
\begin{equation}
  f (\omega_s,\omega_i )=f_s (\omega_s )f_i (\omega_i ).
\end{equation}
Thus, K = 1.

For the spectral correlation case, to calculate the numerical results of the Schimdt mode, we could discretize the spectrum. The JSA can be expressed in a large square matrix $F$. The matrix element $F_{mn}$ of the $m$th row and $n$th column is the discrete frequency value of $f (\omega_{s,m},\omega_{i,n})$, where $\omega_{s,m}$ and $\omega_{i,n}$ are discrete frequencies. The mathematical singular value decomposition (SVD) can work to decompose the matrix $F$ into the following:
\begin{equation}
  F=UDV=U\left[ 
  \begin{matrix}
  	 d_1 &&&\\ 
  	 &d_2&&\\
  	 && \ddots \\
  \end{matrix} 
  \right] V,
\end{equation}
where $U$ and $V$ are unitary matrices, $U$ depends on $\omega_s$, and $V$ depends on $\omega_i$. The diagonal matrix D needs to be normalized with a coefficient $\frac{1}{\sum d_n^2}$. The matrix of the weight of Schmidt mode is then 

\begin{equation}
	\left[ 
  \begin{matrix}
  	 \lambda_1 &&&\\ 
  	 &\lambda_2&&\\
  	 && \ddots \\
  \end{matrix} 
  \right]
  =
  \frac{1}{\sum d_n^2}
  \left[ 
  \begin{matrix}
  	 d_1^2 &&&\\ 
  	 &d_2^2&&\\
  	 && \ddots \\
  \end{matrix} 
  \right].
\end{equation}

\section{HOM interference theory}
HOM interference acts as an essential role in revealing the nonclasscial multiphoton phenomenon. Its visibility is directly related to the purity of the photon source. Down-converted photon pairs can serve as a single-photon source, where one of the photons (idler in this work) serves as a trigger. A HOM interference can be arranged by combining two independent heralded sources on a beam-splitter. The four-fold coincidence probability can be written as follows \cite{ou2007multi} :
\begin{eqnarray}
P (t)=&\int  d\omega_{s_1} d\omega_{s_2} d\omega_{i_1} d\omega_{i_2} \vert f' (\omega_{s_1},\omega_{i_1}) f' (\omega_{s_2},\omega_{i_2})\nonumber \\
&- 
f' (\omega_{s_2},\omega_{i_1}) f' (\omega_{s_1},\omega_{i_2}) \cdot e^{-i (\omega_{s_2} - \omega_{s_1})t} \vert ^2,
\end{eqnarray}
where $t$ denotes the path delay between two independent sources.

When the independent single photons are indistinguishable, the HOM interference generates a dip in four-fold coincidence when the time delay is 0. Also, the visibility is defined as follows:
\begin{equation}
\label{equ:vis}
V=\frac{P (\infty)- P (0)}{P (\infty)}=\mathcal{E}/\mathcal{A},
\end{equation}
\begin{equation}
	\mathcal{A}=\int d\omega_{s_1} d\omega_{s_2} d\omega_{i_1} d\omega_{i_2} \left\vert f^\prime (\omega_{s1},\omega_{i1})f^\prime (\omega_{s2},\omega_{i2}) \right\vert ^2,
\end{equation}
\begin{eqnarray}
\mathcal{E}=&\int d\omega_{s_1} d\omega_{s_2} d\omega_{i_1} d\omega_{i_2} f^\prime (\omega_{s1},\omega_{i1})f^\prime (\omega_{s2},\omega_{i2})\nonumber\\& \cdot f^{\prime*} (\omega_{s1},\omega_{i2})f^{\prime*} (\omega_{s2},\omega_{i1})
\end{eqnarray}

where $f^{\prime*}$ signifies a conjugate function.

Visibility  is theoretically equal to the state purity, $V=P=\frac{1}{K}$ \cite{Mosley2008a}, which is the straightforward method for evaluating spectral purity.

%

\end{document}